\documentclass[12pt]{article}

\usepackage[centertags]{amsmath}
\usepackage{amsmath}
\usepackage{amsfonts}
\usepackage{amssymb}
\usepackage{amsthm}
\usepackage{newlfont}
\usepackage{epsfig}
\usepackage{amscd}

\newcommand{\RR}{{\mathbb R}}

\newcommand{\CC}{{\mathbb C}}
\newcommand{\beq}{\begin{equation}}
\newcommand{\eeq}{\end{equation}}
\newcommand{\ba}{\begin{array}}
\newcommand{\ea}{\end{array}}
\newcommand{\bea}{\begin{eqnarray}}
\newcommand{\eea}{\end{eqnarray}}

\begin{document}

\begin{center}
{\large \sc \bf Solvable vector nonlinear Riemann problems, \\ exact implicit solutions 
of dispersionless PDEs \\ and wave breaking }

\vskip 20pt

{\large  S. V. Manakov$^{1,\S}$ and P. M. Santini$^{2,\S}$}

\vskip 20pt

{\it 
$^1$ Landau Institute for Theoretical Physics, Moscow, Russia

\smallskip

$^2$ Dipartimento di Fisica, Universit\`a di Roma "La Sapienza", and \\
Istituto Nazionale di Fisica Nucleare, Sezione di Roma 1 \\
Piazz.le Aldo Moro 2, I-00185 Roma, Italy}

\bigskip

$^{\S}$e-mail:  {\tt manakov@itp.ac.ru, paolo.santini@roma1.infn.it}

\bigskip

{\today}

\end{center}

\begin{abstract}We have recently solved the inverse spectral problem for integrable PDEs in arbitrary dimensions 
arising as commutation of multidimensional vector fields depending on a spectral parameter $\lambda$. The associated inverse 
problem, in particular, can be formulated as a nonlinear Riemann Hilbert (NRH) problem on a given contour of the 
complex $\lambda$ plane. The most distinguished examples of integrable PDEs of this type, like the dispersionless 
Kadomtsev-Petviashivili (dKP), the heavenly and the 2 dimensional dispersionless Toda equations, are real PDEs  
associated with Hamiltonian vector fields. The corresponding NRH data satisfy suitable reality and symplectic 
constraints. In this paper, generalizing the examples of solvable NRH problems illustrated in \cite{MS4,MS5,MS6},  
we present a general procedure to construct solvable NRH problems for integrable real PDEs associated with 
Hamiltonian vector fields, allowing one to construct implicit solutions of such PDEs parametrized 
by an arbitrary number of real functions of a single variable. Then we  
illustrate this theory on few distinguished examples for the dKP and heavenly equations. For the dKP case, 
we characterize a class of similarity solutions, a class of 
solutions constant on their parabolic wave front and breaking simultaneously on it, and a class of 
localized solutions breaking in a point of the $(x,y)$ plane. For the 
heavenly equation, we characterize two classes of symmetry reductions.

\end{abstract}

\section{Introduction}

It was observed long ago \cite{ZS} that the commutation of multidimensional 
vector fields can generate integrable nonlinear partial differential equations (PDEs) in arbitrary 
dimensions. Some of these equations are dispersionless (or quasi-classical) limits 
of integrable PDEs, having the dispersionless Kadomtsev 
- Petviashvili (dKP) equation \cite{Timman},\cite{ZK} as universal prototype example; 
they arise in various problems of 
Mathematical Physics and are intensively studied in the recent literature 
(see, f.i., \cite{KG} - \cite{KM}). In particular, an elegant integration scheme applicable, in general,  
to nonlinear PDEs associated with Hamiltonian vector fields, was presented in \cite{Kri2} and a 
nonlinear $\bar\partial$ - dressing was developed in \cite{K-MA-R}. Special classes of nontrivial solutions 
were also derived (see, f.i., \cite{DT}, \cite{G-M-MA}).

Distinguished examples of PDEs arising as the commutation conditions $[\hat L_1(\lambda),\hat L_2(\lambda)]=0$ of pairs of one 
parameter families of vector fields, being $\lambda\in\CC$ the spectral parameter, are the following. \\
1. The vector nonlinear PDE in $N+4$ dimensions \cite{MS1}: 
\beq
\label{quasilin-U}
\vec U_{t_1z_2}-\vec U_{t_2z_1}+
\left(\vec U_{z_1}\cdot\nabla_{\vec x}\right)\vec U_{z_2}-\left(\vec U_{z_2}\cdot\nabla_{\vec x}\right)\vec U_{z_1}=\vec 0,
\eeq
where $\vec U(t_1,t_2,z_1,z_2,\vec x)\in\RR^N$, $\vec x=(x_1,\dots,x_N)\in\RR^N$ and  
$\nabla_{\vec x}=(\partial_{x_1},..,\partial_{x_N})$, associated with the following pair of $(N+1)$ dimensional 
vector fields
\beq
\label{L1L2_quasilin-U}
\hat L_i=\partial_{t_i}+\lambda\partial_{z_i}+\vec U_{z_i}\cdot\nabla_{\vec x},~~~~i=1,2. 
\eeq  
2. Its dimensional reduction, for $N=2$ \cite{MS1}:
\beq
\label{quasilin-U-bis}
\ba{l}
\vec U_{tx}-\vec U_{zy}+\left(\vec U_{y}\cdot\nabla_{\vec x}\right)\vec U_{x}-
\left(\vec U_{x}\cdot\nabla_{\vec x}\right)\vec U_{y}=\vec 0, \\
\vec U\in\RR^2,~~\vec x=(x,y),~~\nabla_{\vec x}=(\partial_x,\partial_y),
\ea
\eeq
obtained renaming the independent variables as follows: $t_1=z,~t_2=t,~~x_1=x,~~x_2=y$, associated with the 
two-dimensional vector fields:
\beq
\label{L1L2_quasilin-U-bis}
\ba{l}
\hat L_1=\partial_{z}+\lambda\partial_{x}+\vec U_x\cdot\nabla_{\vec x},     \\
\hat L_2=\partial_{t}+\lambda\partial_{y}+\vec U_y\cdot\nabla_{\vec x}.
\ea
\eeq
3. The Hamiltonian reduction $\nabla_{\vec x}\cdot\vec U=0$ of (\ref{quasilin-U-bis}), the celebrated second 
heavenly equation of Plebanski \cite{Pleb}: 
\beq
\label{heavenly}
\theta_{tx}-\theta_{zy}+\theta_{xx}\theta_{yy}-\theta^2_{xy}=0,~~~
\theta=\theta(x,y,z,t)\in\RR,~~~~~x,y,z,t\in\RR,
\eeq
describing self-dual vacuum solutions of the Einstein equations, associated with 
the following pair of Hamiltonian two-dimensional vector fields  
\beq
\label{L1L2_heav}
\ba{l}
\hat L_1\equiv \partial_{z}+\lambda\partial_{x}+\theta_{xy}\partial_{x}-\theta_{xx}\partial_{y}, \\
\hat L_2\equiv \partial_{t}+\lambda\partial_{y}+\theta_{yy}\partial_{x}-\theta_{xy}\partial_{y}. 
\ea
\eeq
4. The following system of two nonlinear PDEs in $2+1$ dimensions \cite{MS2}:
\beq
\label{dKP-system}
\ba{l}
u_{xt}+u_{yy}+(uu_x)_x+v_xu_{xy}-v_yu_{xx}=0,         \\
v_{xt}+v_{yy}+uv_{xx}+v_xv_{xy}-v_yv_{xx}=0,
\ea
\eeq
arising from the commutation of the two-dimensional vector fields 
\beq\label{Ltilde}
\ba{l}
\tilde L_1\equiv \partial_y+(\lambda+v_x)\partial_x-u_x\partial_{\lambda}, \\
\tilde L_2\equiv \partial_t+(\lambda^2+\lambda v_x+u-v_y)\partial_x+(-\lambda u_x+u_y)\partial_{\lambda}, 
\ea
\eeq
and describing a general integrable Einstein-Weyl metric \cite{Dunajj}.                       \\
5. The $v=0$ reduction of (\ref{dKP-system}), the dKP equation  
\beq
\label{dKP}
(u_t+uu_x)_x+u_{yy}=0,~~~
u=u(x,y,t)\in\RR,~~~~~x,y,t\in\RR,
\eeq 
(the $x$-dispersionless limit of the celebrated Kadomtsev-Petviashvili equation \cite{KP}),  
associated with the following pair of Hamiltonian two-dimensional vector fields \cite{Zakharov,Kri2}:
\beq
\label{L1L2_dKP}
\ba{l}
\hat L_1\equiv \partial_y+\lambda\partial_x-u_x\partial_{\lambda},    \\
\hat L_2\equiv \partial_t+(\lambda^2+u)\partial_x+(-\lambda u_x+u_y)\partial_{\lambda},
\ea
\eeq
describing the evolution of small amplitude, nearly one-dimensional waves in shallow water \cite{AC} 
near the shore (when the $x$-dispersion can be neglected), as well as unsteady motion in transonic flow 
\cite{Timman} and nonlinear acoustics of confined beams \cite{ZK}.     \\ 
6. The $u=0$ reduction of (\ref{dKP-system}) \cite{MA-S,Pavlov}:
\beq 
\label{Pavlov}
\ba{l}
v_{xt}+v_{yy}=v_yv_{xx}-v_xv_{xy},~~~~~~~ 
v=v(x,y,t)\in\RR,~~~~~x,y,t\in\RR,
\ea
\eeq
associated with the non-Hamiltonian one-dimensional vector fields \cite{Duna}
\beq
\label{L1L2_Pavlov}
\ba{l}
\hat L_1\equiv \partial_y+(\lambda +v_x)\partial_x, \\
\hat L_2\equiv \partial_t+(\lambda^2+\lambda v_x-v_y)\partial_x.
\ea
\eeq
7. The two-dimensional dispersionless Toda (2ddT) equation \cite{FP,Zak}:
\beq\label{2ddT}
\phi_{\zeta_1\zeta_2}=\left(e^{\phi_t}\right)_t,~~~\phi=\phi(\zeta_1,\zeta_2,t)   
\eeq
(or $\varphi_{\zeta_1\zeta_2}=\left(e^{\varphi}\right)_{tt},~\varphi=\phi_t$), associated with the pair of 
Hamiltonian vector fields \cite{TT1}:    
\beq\label{L1L2_2ddT}
\ba{l}
\hat L_1=\partial_{\zeta_1}+\lambda e^{\frac{\phi_{t}}{2}}\partial_{t}+
\left(-\lambda (e^{\frac{\phi_{t}}{2}})_t+\frac{\phi_{\zeta_1 t}}{2}\right)\lambda\partial_{\lambda} ,\\ 
\hat L_2=\partial_{\zeta_2}+\lambda^{-1} e^{\frac{\phi_{t}}{2}}\partial_{t}+
\left(\lambda^{-1}(e^{\frac{\phi_{t}}{2}})_t-\frac{\phi_{\zeta_2 t}}{2}\right)\lambda\partial_{\lambda}, 
\ea
\eeq  
describing integrable heavens \cite{BF,GD} and Einstein - Weyl geometries \cite{H}, \cite{J}, \cite{Ward};     
whose string equations solutions \cite{TT3} are relevant in the ideal Hele-Shaw problem 
\cite{MWZ}-\cite{MAM}.

The Inverse Spectral Transform (IST) for $1$-parameter families of multidimensional vector fields,  
developed in \cite{MS1},  
has allowed one to construct the formal solution of the Cauchy problem for the nonlinear PDEs 
(\ref{quasilin-U-bis}) and (\ref{heavenly}) in \cite{MS1}, for equations (\ref{dKP-system}) and (\ref{dKP}) 
in \cite{MS2}, for equation (\ref{Pavlov}) in \cite{MS3} and for the wave form 
$(e^{\phi_t})_t=\phi_{xx}+\phi_{yy}$ of equation (\ref{2ddT}) in \cite{MS4}.    
This IST, introducing interesting novelties 
with respect to the classical IST for soliton equations \cite{ZMNP,AC}, turns out to be, together with its associated
 nonlinear Riemann - Hilbert (NRH) dressing, an efficient 
tool to study several properties of the solution space of the PDE under consideration: 
i) the characterization of a 
distinguished class of spectral data for which the associated nonlinear RH problem is linearized and solved, 
corresponding to a class of implicit solutions of the PDE (for the dKP and 2ddT equations respectively 
in \cite{MS5} and in \cite{MS4}, and for the Dunajski generalization \cite{Duna2} of the heavenly equation in \cite{BDM}) 
and for equations (\ref{Pavlov}) and (\ref{heavenly}) in \cite{MS6};  
ii) the construction of the longtime behaviour of the solutions of the Cauchy problem (for the dKP,  
2ddT and heavenly equations respectively in \cite{MS6}, \cite{MS5} and \cite{MS6});     
iii) the possibility  to establish  whether or not the lack of dispersive terms in the nonlinear PDE 
causes the breaking of localized initial profiles (for the dKP, 2ddT and heavenly equations respectively 
in \cite{MS5}, in \cite{MS4} and in \cite{MS6}) and, if yes, to investigate in a surprisingly 
explicit way the analytic aspects of such a wave breaking, as it was done for the dKP equation  
in \cite{MS5}. Recent results on integrable differential constraints on the hierarchy associated 
with the nonlinear system (\ref{dKP-system}) and their 
connection to the associated NRH problems can be found in \cite{Bogdanov}.

In this paper, generalizing the examples of solvable NRH problems illustrated in \cite{MS4,MS5,MS6},  
we present, in \S 2, a general procedure to construct solvable NRH problems for integrable PDEs associated with 
Hamiltonian vector fields, allowing one to construct implicit solutions of such PDEs parametrized 
by an arbitrary number of real functions of a single variable. In \S 3 we  
illustrate this theory on few distinguished examples for the dKP equation, including the similarity solutions, a class of 
solutions constant on their parabolic wave front and breaking simultaneously on it, and a class of 
localized solutions breaking in a point of the $(x,y)$ plane. In \S 4 we briefly apply the theory to 
the heavenly equation, constructing their similarity solutions.

Since the theory will be illustrated on two basic examples of PDEs associated with Hamiltonian vector fields: 
the dKP and the heavenly equations, in the remaining part of this introductory section we summarize their 
NRH dressing formalisms.   

\subsection{NRH dressing for dKP \cite{MS2},\cite{MS5}} 

Consider the vector nonlinear RH problem on the real line:  
\beq\label{RH_dKP}
\ba{l}
\psi^{+}_1={\cal R}_1\left(\psi^{-}_1,\psi^{-}_2\right), \\
\psi^{+}_2={\cal R}_2\left(\psi^{-}_1,\psi^{-}_2\right),~~\lambda\in\RR,
\ea
\eeq
or, more shortly, 
\beq\label{vectorRH}
\vec\psi^+(\lambda)=\vec{\cal R}(\vec\psi^-(\lambda)),
\eeq
where ${\cal R}_j(s_1,s_2),~j=1,2$ is a given pair of complex differentiable functions of two 
arguments, and the  
solutions $\vec\psi^{\pm}(\lambda)=(\psi^{\pm}_1(\lambda),\psi^{\pm}_2(\lambda))\in\CC^2$ 
are $2$-dimensional vector functions analytic respectively in the upper and lower 
halves of the complex $\lambda$ plane, with the following asymptotics, for $|\lambda |\gg 1$ 
in their analyticity domains:
\beq
\label{psi_asympt_1}
\ba{l}
\psi^{\pm}_1(\lambda)=-\lambda^2t-\lambda y+x-2tq^{(1)}_2+\sum\limits_{n\ge 1}\frac{q^{(n)}_1}{\lambda},\\
\psi^{\pm}_2(\lambda)=\lambda+\frac{q^{(1)}_2}{\lambda}+\sum\limits_{n\ge 2}\frac{q^{(n)}_2}{\lambda},
\ea
\eeq
where, f.i.:
\beq
\ba{l}
\label{psi_asympt_2}
q^{(1)}_1=-yu+2t\partial^{-1}_xu_{y},~~q^{(2)}_1=\partial^{-1}_x(yu)_y-2t\partial^{-2}_xu_{yy}, \\
q^{(1)}_2=u,~~q^{(2)}_2=-\partial^{-1}_xu_y,~~q^{(3)}_2=-u^2/2+\partial^{-2}_xu_{yy}, \\
q^{(4)}_2=\frac{1}{2}\partial^{-1}_x(u^2)_y+2\partial^{-1}_x(u_x\partial^{-1}_xu_y)-\partial^{-3}_xu_{yyy}.
\ea
\eeq
Then, if the nonlinear RH problem (\ref{RH_dKP}) is uniquely solvable, together with its 
linearized version 
\beq
\ba{l}
\vec v^{+}=M\left(\psi^{-}_1,\psi^{-}_2\right) \vec v^{-}, \\
M_{jk}\left(\psi^{-}_1,\psi^{-}_2\right)=
\frac{\partial{\cal R}_j\left(\psi^{-}_1,\psi^{-}_2\right)}{\partial \psi^{-}_k},~~j,k=1,2,
\ea
\eeq
then $\psi^{\pm}_{1,2}$ are solutions of the following pair of vector field equations:
\beq
\ba{l}
\hat L_1\psi\equiv \psi_y+(\lambda+v_x)\psi_x-u_x \psi_{\lambda}=0, \\
\hat L_2\psi\equiv \psi_t+(\lambda^2+\lambda v_x+u-v_y)\psi_x+(-\lambda u_x+u_y)\psi_{\lambda}=0,
\ea
\eeq
where
\beq
\ba{l}
u=q^{(1)}_2, \\
v=-q^{(1)}_1-y q^{(1)}_2-2 t q^{(2)}_2,
\ea
\eeq
and $u$ and $v$ solve the system (\ref{dKP-system}) of PDES. \\
\\ 
2. {\bf Basic non-differential reductions}. There are few basic non-differential reductions of the above NRH 
problem and, correspondingly, of the integrable system (\ref{dKP-system}). \\
\\
{\bf R1}. If ${\cal R}_2(s_1,s_2)=s_2$, then $\psi^+_2=\psi^-_2=\lambda$, the nonlinear 
RH problem becomes scalar and one obtains equation (\ref{Pavlov}), the $u=0$ reduction of the system (\ref{dKP-system}),  
associated with the non Hamiltonian Lax pair of one - dimensional vector fields (\ref{L1L2_Pavlov}). \\
\\
{\bf R2}. If the transformation $(s_1,s_2)\to ({\cal R}_1(s_1,s_2),{\cal R}_2(s_1,s_2))$ is canonical:
\beq\label{R-dKP}
\{{\cal R}_1,{\cal R}_2 \}_{(s_1,s_2)}:={{\cal R}_1}_{s_1}{{\cal R}_2}_{s_2}-{{\cal R}_1}_{s_2}{{\cal R}_2}_{s_1}
=1,
\eeq  
one obtains the $v=0$ reduction of the system (\ref{dKP-system}), 
the celebrated dKP equation (\ref{dKP}), corresponding to the Hamiltonian Lax pair
\beq\label{Lax1dKP}
\ba{l}
\psi_y+\lambda\psi_x-u_x \psi_{\lambda}=\psi_y+\{H_2,\psi\}_{(\lambda,x)}=0, 
\ea
\eeq
\beq\label{Lax2dKP}
\ba{l}
\psi_t+(\lambda^2+u)\psi_x+(-\lambda u_x+u_y)\psi_{\lambda}=\psi_t+\{H_3,\psi\}_{(\lambda,x)}=0.
\ea
\eeq
for the Hamiltonians
\beq
H_2=\frac{\lambda^2}{2}+u,~~ H_3=\frac{\lambda^3}{3}+\lambda u-\partial^{-1}_xu_y .
\eeq
We remark that, in this reduction, the eigenfunctions $\psi^{\pm}_1,\psi^{\pm}_2$ are 
canonically conjugated: 
\beq
\{\psi^{\pm}_2,\psi^{\pm}_1\}_{(\lambda,x)}=1
\eeq
We also recall that \cite{TT1,Kri1} the dKP equation is the first member of the dKP hierarchy
\beq\label{hierarchy}
\ba{l}
{H_m}_{t_n}-{H_n}_{t_m}+\{H_n,H_m\}_{(\lambda,x)}=0,~~m\ne n,~~m,n\ge 2, \\
H_n\equiv \frac{1}{n}\left(({\psi^+_2})^n\right)_+,~~n\ge 2,
\ea
\eeq
corresponding to the Hamiltonian vector field equations
\beq
\psi_{t_n}+\{H_n,\psi \}_{(\lambda,x)}=0,
\eeq
where $t_2=y$, $t_3=t$ and $(\psi )_+$ is the non negative (principal) part of the Laurent expansion of $\psi$ 
at $\lambda=\infty$. In particular, if $m=2$ in (\ref{hierarchy}), one obtains the sub-hierarchy 
\beq\label{sub_hierarchy}
nu_{t_n}+\left({\mbox{Res}}({\psi^+_2}^n)\right)_x=0.
\eeq
where Res$(g)$ is the coefficient of $\lambda ^{-1}$ in the Laurent expansion of $g(\lambda)$ at $\lambda=\infty$.\\
\\
{\bf R3}. If 
\beq\label{reality1}
\vec{\cal R}(\overline{ \vec{\cal R}(\bar{\vec\zeta})})=\vec\zeta,
~~~\forall\vec\zeta\in\CC^2,
\eeq
then one obtains the reality constraint
\beq\label{reality2}
u,v\in \RR,~~\psi^{+}_j(\lambda)=\overline{\psi^{-}_j(\overline{\lambda})},~j=1,2.
\eeq

From the integral equations  
characterizing the solutions of the NRH problem (\ref{RH_dKP}), and from the definition $u=q^{(1)}_2$,   
one obtains the following spectral characterization of the solution $u$:
\beq\label{inverse_1}
u=F\left(x-2ut,y,t\right)\in\RR,
\eeq
where the spectral function $F$, defined by 
\beq\label{inverse_2}
\ba{l}
F\left(\xi,y,t \right)=-\int\limits_{\RR}\frac{d\lambda}{2\pi i}R_2
\Big(\pi^-_1(\lambda;\xi,y,t),\pi^-_2(\lambda;\xi,y,t)\Big),  \\
{\cal R}_j(\zeta_1,\zeta_2)=\zeta_j+R_j(\zeta_1,\zeta_2),~~~j=1,2,
\ea
\eeq
is connected with the initial data $u(x,y,0)$ via the direct spectral transform 
developed in \cite{MS2}. Equation (\ref{inverse_1}) defines the dKP solution implicitly , 
due to the presence of the $(x-2ut)$ term as argument of $F$, and describes the 
wave breaking features of localized solutions of dKP. 

Evaluating the integral equations characterizing the above NRH problem in the 
space-time asymptotic region
\beq\label{parabola}
\ba{l}
\xi -2ut,v_1,v_2=O(1),~~v_2\ne 0,~~t\gg 1, \\
x+\frac{y^2}{4t}=\xi,~~x=\xi +v_1t,~~y=v_2t ,
\ea
\eeq
one obtains the following longtime behavior \cite{MS5}
\beq
\label{asympt_dKP_1}
u= \frac{1}{\sqrt{t}}G\left(x+\frac{y^2}{4t}-2ut,\frac{y}{2t}\right)+
o\left(\frac{1}{\sqrt{t}}\right), 
\eeq
where
\beq
\label{asympt_dKP_2}
G(\xi,\eta)=
-\frac{1}{2\pi i}\int\limits_{\RR}d\mu
R_2\Big(\xi-\mu^2+a_1(\mu;\xi,\eta),-\eta+a_2(\mu;\xi,\eta) \Big),
\eeq
and $a_{1,2}(\mu;\xi,\eta)$ are the unique solutions of the nonlinear integral equations
\beq\label{a_dKP}
\ba{l}
a_j(\mu;\xi,\eta)= 
\frac{1}{2\pi i}\int\limits_{\RR}\frac{d\mu'}{\mu'-(\mu-i0)}
R_j\Big(\xi-{\mu'}^2+a_1(\mu';\xi,\eta),   \\
-\eta+a_2(\mu';\xi,\eta) \Big),~~~~~j=1,2 ,
\ea
\eeq
giving a description of the wave breaking of small initial data in the longtime regime as 
explicit as for the case of the Riemann Hopf equation $u_t+uu_x=0$. In particular, one shows 
that small and localized initial data will break asympotically in a point of the $(x,y)$ 
plane and the breaking details are given explicitly in terms of the initial data \cite{MS5},\cite{MS7}. 
 
\subsection{NRH dressing for heavenly \cite{MS1,MS6}}

Consider the vector NRH problem on the real line:  
\beq\label{RH_heav}
\vec\psi^+(\lambda)=\vec\psi^-(\lambda)+\vec R(\vec\psi^-(\lambda),\lambda),~~\lambda\in\RR,
\eeq
where $\vec\psi^{+}(\lambda),\vec\psi^{-}(\lambda)\in\CC^2$ are $2$-dimensional 
vector functions analytic in the upper and lower 
halves of the complex $\lambda$ plane, normalized as follows 
\beq\label{normal_heav}
\ba{l}
\vec\psi^{\pm}(\lambda)=\left(
\ba{c}
x-\lambda z \\
y-\lambda t
\ea
\right)
+O(\lambda^{-1}),~~|\lambda |\gg 1, 
\ea
\eeq 
and the spectral data $\vec R(\vec \zeta,\lambda)=
(R_1(\zeta_1,\zeta_2,\lambda),R_2(\zeta_1,\zeta_2,\lambda))\in\CC^2$, defined for 
$\vec \zeta\in\CC^2,~~\lambda\in\RR$, satisfy the following properties:
\beq\label{R_heav}
\ba{ll}
\vec{\cal R}(\overline{ \vec{\cal R}(\bar{\vec\zeta},\lambda)},\lambda)=\vec\zeta,
~~~\forall\vec\zeta\in\CC^2,  &  \mbox{reality constraint,} \\
\{{\cal R}_1,{\cal R}_2\}_{\vec\zeta}=1,  &  \mbox{symplectic constraint},
\ea
\eeq
where $\vec{\cal R}(\vec\zeta,\lambda):=\vec\zeta+\vec R(\vec\zeta,\lambda)$. Then, assuming uniqueness of the solution 
of such a NRH problem and of its linearized version, it follows that $\vec\psi^{\pm}$ are 
canonically conjugated solutions ($\{\psi^{\pm}_1,\psi^{\pm}_2\}_{(x,y)}=1$) of the linear problems 
$\hat L_{1,2}\vec\psi^{\pm}=\vec 0$, where $\hat L_{1,2}$ are defined in (\ref{L1L2_heav}), and   
\beq\label{pot_1_heav}
\left(
\ba{c}
\theta_y \\
-\theta_x
\ea
\right)=\vec F(x,y,z,t) \in\RR^2
\eeq
is solution of the heavenly equation (\ref{heavenly}), where
\beq\label{pot_2_heav}  
\vec F(x,y,z,t)\equiv \int\limits_{\RR}\frac{d\lambda}{2\pi i}
\vec R\Big(\psi^-_1(\lambda;x,y,z,t),\psi^-_2(\lambda;x,y,z,t),\lambda \Big).
\eeq
As a consequence of equations $\hat L_{1,2}\vec\psi^{\pm}=\vec 0$, it follows that, for $|\lambda |\gg 1$: 
\beq\label{psi_asympt_heav}
\ba{l}
\psi^{\pm}_1=x-\lambda z-\theta_y \lambda^{-1}+\theta_t \lambda^{-2} +O(\lambda^{-3}), \\
\psi^{\pm}_2=y-\lambda t+\theta_x \lambda^{-1}-\theta_z \lambda^{-2} +O(\lambda^{-3}).
\ea
\eeq

\section{Solvable vector nonlinear RH problems} 
Since basic nonlinear PDEs like the dKP, heavenly and 2ddT equations are solved via NRH problems 
whose data satisfy the symplectic and reality constraints,  
in this section we present a general procedure to construct solvable vector NRH problems satisfying 
such constraints.

Consider an authonomous Hamiltonian two-dimensional dynamical system with Hamiltonian 
\beq\label{Hamiltonian}
H(\underline x)={\cal H}(S(\underline x))
\eeq
where $\underline x\equiv (q,p)$ are canonically conjugated coordinates, ${\cal H}(\cdot )$ 
is an arbitrary entire function of a single argument and $S(\underline x)$ is an entire function 
of the coordinates, corresponding to the equations of motion
\beq\label{dyn_syst1}
\frac{d\underline x}{dt}={\cal H}'(S)
\left(\ba{cc}
0 & 1 \\
-1 & 0
\ea\right)
\nabla_{\underline x}S(\underline x). 
\eeq

Introducing action - angle variables in the usual way:
\beq\label{action-angle}
\ba{l}
J\equiv \frac{1}{2\pi}\oint p(q,H)dq,~~\Rightarrow~~H=H(J), \\
\theta-\theta_0\equiv \omega(J){\cal H}'(S)(t-t_0)={\cal H}'(S)\int\limits_{q_0}^q \frac{\partial p(q',H(J))}{\partial J}dq' ,\\
\omega(J)\equiv \frac{\partial H(J)}{\partial J} ,
\ea
\eeq
the solution can be found inverting the quadrature (\ref{action-angle}):
\beq\label{sol_x}
\vec x=\vec {\cal D}(\theta-\theta_0;\vec x_0,J)
\eeq
where the evolutionary diffeomorphism $\vec {\cal D}:~\vec x_0\to\vec x$ is symplectic:
\beq
\{{\cal D}_1,{\cal D}_2\}_{(q_0,p_0)}=1
\eeq
and satisfies the properties:
\beq\label{properties-R}
\ba{l}
\vec x_0=\vec {\cal D}(\theta_0-\theta;\vec x,J), \\
\vec {\cal D}(\theta_2;\vec {\cal D}(\theta_1;\vec x_0,J),J)=\vec {\cal D}(\theta_1+\theta_2;\vec x_0,J), \\
\vec {\cal D}(0;\vec x_0,J)=\vec x_0 .
\ea
\eeq 
Consequently, we have the relation
\beq\label{sol-RH1}
\vec {\cal D}(\theta_1-\theta;\vec x,J)=\vec {\cal D}(\theta_1-\theta_0;\vec x_0,J)
\eeq
for any intermediate angle $\theta_1$, and this relation holds true, at least locally, also 
in the complex domain. 

\subsection{Solvable NRH problems}
Identifying the solution $\vec x(t)$ at time $t_0$ and at time $t_0+1$ with, respectively, $\vec \psi^-(\lambda)$ and 
$\vec\psi^+(\lambda)$, \\
i) equation (\ref{sol_x}) becomes the two-dimensional vector NRH problem 
\beq\label{NRH4}
\vec\psi^+=\vec {\cal D}\left(\omega(J(\vec \psi^-)){\cal H}'(S(\vec \psi^-));\vec \psi^-,J(\vec \psi^-)\right)\equiv 
\vec {\cal R}\left(\vec \psi^-\right)
\eeq
connecting the $(-)$ and $(+)$ vector functions through the canonical transformation (\ref{R-dKP}); \\
ii) since $S(\underline x)$ (as well as $J(\vec x)$) is an invariant of the dynamics: $S(\vec x_0 )=S(\vec x)$, 
it follows that 
$S(\vec \psi^-)=S(\vec \psi^+)$ is an invariant of the NRH problem (\ref{NRH4}). Therefore 
$S(\vec \psi^{\pm})$ define an entire function, given by their polynomial part $W(\lambda)$:
\beq\label{invariance}
S(\vec \psi^-(\lambda))=S(\vec \psi^+(\lambda))\equiv W(\lambda). 
\eeq 
In addition, since $S$ is a real function of its arguments, the reality constraint (\ref{reality2}) implies that
\beq\label{reality_W}
\overline{W(\bar\lambda)}=W(\lambda).
\eeq 
iii) Since also $\vec {\cal D}$ is a real function of its arguments and, from  (\ref{properties-R}),
\beq
\vec {\cal D}(z;\vec {\cal D}(-z;\vec \zeta,J),J)=\vec \zeta,
\eeq
the reality constraint (\ref{reality1}) implies that $\vec {\cal D}(-z;\vec\zeta,J)=\vec {\cal D}(\bar z;\vec\zeta,J)$;  
it follows, from (\ref{NRH}), that 
\beq\label{reality3}
{\cal H'}(\cdot)=if(\cdot) ,
\eeq
where $f$ is an arbitrary real function of a single argument, for dKP, and of two arguments (the invariant (\ref{invariance}) 
and $\lambda$),  for heavenly. \\ 
iv) Defining the upper and lower functions
\beq
\theta^{\pm}(\lambda )\equiv \frac{1}{2\pi i}\int\limits_{\RR}\frac{d\lambda'}{\lambda'-
(\lambda\pm i0)}\left(i\omega(J(\vec \psi^-(\lambda')))f(W(\lambda'))\right)(\lambda'),
\eeq
so that $i\omega f=\theta^+-\theta^-$, becomes
\beq\label{sol-RH2}
\vec {\cal D}(-\theta^+;\psi^+,J(\psi^+))=\vec {\cal D}(-\theta^-;\psi^-,J(\psi^-)),
\eeq
and provides the solution of the NRH problem, at least if $\vec {\cal D}$ is an entire function of its arguments.  
Indeed, in this case, the left and right hand sides of (\ref{sol-RH2}) 
are equal to their polynomial part $\vec E(\lambda)$ at $\lambda\sim\infty$:
\beq\label{sol-RH3}
\vec {\cal D}(-\theta^{\pm};\vec\psi^{\pm},J(\vec\psi^{\pm}))=\vec E(\lambda)
\eeq
and these are algebraic equations for $\vec \psi^{\pm}(\lambda)$. 

\vskip 5pt
\noindent
Remark 1. Solvable NRHs of this type allow one to construct solutions of the dispersionless PDE 
characterized by differential reductions of the integrable PDE. 
To show it, let us consider, for concreteness, the dKP case. Let $g(\lambda)$ be a solution of the NRH problem,  
eigenfunction of the spectral problem (\ref{Lax1dKP}), and characterized by the formal Laurent expansion:
\beq
g(\lambda)\sim \sum\limits_{n=-\infty}^N g_n\lambda^n,~~|\lambda|\gg 1.
\eeq
Then the coefficients $g_n$ of $\lambda^n$ satisfy the recursion relations
\beq\label{alpha_equ}
\ba{l}
{g_{N}}_x=0,~~{g_{N}}_y+{g_{N-1}}_x=0, \\
{g_{n}}_y+{g_{n-1}}_x-(n+1)u_x g_{n+1}=0,~~n<N.
\ea
\eeq
Since $W(\lambda)$, defined by (\ref{invariance}), is a polynomial (in $\lambda$) solution of  
(\ref{Lax1dKP}), then Res$_{\infty}W(\lambda)\equiv W_{-1}=0$ and equation (\ref{alpha_equ}) for $n=0$ yields:
\beq\label{diff_constr}
{W_0}_y-u_x{W_1}=-{W_{-1}}_x=0.
\eeq 
In addition, equations (\ref{alpha_equ}) for $n<0$ 
imply, as it has to be, that ${W}_n=0$ for $n\le -1$. Equation  (\ref{diff_constr}) is the differential constraint satisfied 
by the solutions of dKP associated with the solvable NRH problem. 
\vskip 5pt
\noindent
Remark 2. Starting with any $2n$ - dimensional Liouville integrable Hamiltonian 
system, a generalization of the above construction allows one to obtain a solvable $2n$-dimensional NRH problem 
$\vec\psi^+(\lambda)=\vec{\cal R}(\vec\psi^-(\lambda))$, where $\vec{\cal R}:~\vec\psi^-\to \vec\psi^+$ is a 
symplectic transformation.  
\vskip 5pt
\noindent
Remark 3. Also integrable symplectic maps from $\RR^n$ to $\RR^n$ generate, following the same procedure, 
solvable $n$ dimensional vector NRH problems. 

\subsection{Increasing the richness of the solution space}

In the previous section we showed how to construct solvable NRH problems whose spectral data,   
satisfying the reality and symplectic constraints,  have the freedhom of an arbitrary 
real function of a single variable, which parametrizes the corresponding space of implicit solutions of the dKP, 
heavenly and 2ddT equations. In this section we show how  
it is possible to increase considerably the richness of the space of implicit solutions. For 
concreteness we refer to the dKP equation, but the same considerations apply to all real PDEs associated with 
Hamiltonian vector fields.

Let $\psi^{\pm}_{1,2}$ be solutions of the Hamiltonian vector field equations (\ref{Lax1dKP}),(\ref{Lax2dKP}),  
satisfying the asymptotics (\ref{psi_asympt_1}), then arbitrary 
differentiable functions of $(\psi^{\pm}_{1},\psi^{\pm}_{2})$ are also solutions of (\ref{Lax1dKP}),(\ref{Lax2dKP}). 
In the first step of our construction we define new functions of $(\psi^{\pm}_{1},\psi^{\pm}_{2})$ as 
follows:
\beq\label{step1}
\ba{l}
\psi^{(1)\pm}_1=\psi^{\pm}_1+a_1f^{(1)}\left(\psi^{\pm}_2\right), \\
\psi^{(1)\pm}_2=\psi^{\pm}_2
\ea
\eeq
where $a_1\in\RR$ and $f^{(1)}$ is an arbitrary real function of one variable. The transformation (\ref{step1}) 
is clearly invertible and, most important for our purposes, is symplectic and preserves the reality constraint 
(\ref{reality2}). Therefore $\psi^{(1)\pm}_{1,2}$ are also canonically conjugated solutions of the vector field equations 
(\ref{Lax1dKP}),(\ref{Lax2dKP}) satisfying the reality constraint: 
$\{\psi^{(1)\pm}_1,\psi^{(1)\pm}_2\}=1,~\psi^{(1)+}_j(\lambda)=\overline{\psi^{(1)-}_j(\overline{\lambda})},~j=1,2$, 
parametrized by the arbitrary real function $f^{(1)}$ of a single argument. In the second step of this construction,
 we use the solutions $\psi^{(1)\pm}_{1,2}$ to construct new functions $\psi^{(2)\pm}_{1,2}$ as follows  
\beq\label{step2}
\ba{l}
\psi^{(2)\pm}_1=\psi^{(1)\pm}_1, \\
\psi^{(2)\pm}_2=\psi^{(1)\pm}_2+a_2f^{(2)}\left(\psi^{(1)\pm}_1\right)
\ea
\eeq 
where $a_2\in\RR$ and $f^{(2)}$ is another arbitrary real function of one variable 
(notice that, while in (\ref{step1}) the transformation is identical wrt the second eigenfunction, 
in (\ref{step2}) the transformation is identical wrt the first one). Therefore we have constructed the family 
$\psi^{(2)\pm}_{1,2}$ of canonically conjugated solutions of the vector field equations 
(\ref{Lax1dKP}),(\ref{Lax2dKP}) satisfying the reality constraint, and now  parametrized by the two arbitrary real 
functions $f^{(1)},f^{(2)}$ of a single argument. This procedure goes on, without limitations, 
alternating transformations of the type (\ref{step1}) and (\ref{step2}), which are identical respectively 
wrt the second and first eigenfunctions. At the $m^{th}$ step, one constructs canonically conjugated solutions 
$\psi^{(m)\pm}_{1,2}$ of (\ref{Lax1dKP}),(\ref{Lax2dKP}) satisfying the reality constraint:
\beq
\{\psi^{(m)\pm}_1,\psi^{(m)\pm}_2\}=1,~\psi^{(m)+}_j(\lambda)=\overline{\psi^{(m)-}_j(\overline{\lambda})},~j=1,2
\eeq
and parametrized by $m$ arbitrary real functions $f^{(1)},\dots,f^{(m)}$ of a single argument.

Now we consider any of the solvable NRH problems, for the family of eigenfunctions 
$\psi^{(m)\pm}_{1,2}$; for instance, the NRH problem discussed in \S 3.2:
\beq\label{RH_psi^n}
\ba{l}
\psi^{(m)+}_1=\psi^{(m)-}_1+iaf(\psi^{(m)-}_1+a\psi^{(m)-}_2),~~\\
\psi^{(m)+}_2=\psi^{(m)-}_2- if(\psi^{(m)-}_1+a\psi^{(m)-}_2)
\ea
\eeq 
whose normalization easily follows from the definition of $\psi^{(m)\pm}_{1,2}$ in terms of $\psi^{\pm}_{1,2}$, and 
from the asymptotics (\ref{psi_asympt_1}). Then the solution of this exactly solvable NRH problem and 
the corresponding family of exact implicit solutions of dKP are parametrized by $m+1$ arbitrary real 
functions $f,f^{(1)},\dots,f^{(m)}$ of a single argument.

\section{Examples for dKP}
In this section we consider some basic examples of solvable NRH problems associated with dKP.

\subsection{Example 1. The invariant $\psi^+_1 \psi^+_2$ and the similarity reduction}
If $S(q,p)=q p$, then equation (\ref{sol_x}) reads 
\beq
q(t)=q_0 e^{{\cal H'}(S)(t-t_0)},~~p(t)=p_0 e^{-{\cal H'}(S)(t-t_0)};
\eeq 
and corresponds to the NRH problem
\beq
\psi^+_1=\psi^-_1e^{if(\psi^-_1\psi^-_2)},~~\psi^+_2=\psi^-_1e^{-if(\psi^-_1\psi^-_2)},~~\lambda\in\RR ,
\eeq
satisfying the symplectic and reality constraints. Then the invariance equation (\ref{invariance}) becomes:
\beq\label{W_prod}
\psi^+_1 \psi^+_2=\psi^-_1 \psi^-_2=-t\lambda^3-y\lambda^2+(x-3ut)\lambda -2y u+3t\;\partial^{-1}_xu_y \equiv W(\lambda),
\eeq
and the NRH problem linearizes and decouples:
\beq
\ba{l}
\psi^+_1=\psi^-_1e^{if(W(\lambda))},~~
\psi^+_2=\psi^-_2e^{-if(W(\lambda))}.
\ea
\eeq
Equations (\ref{sol-RH3}) become 
\beq
\ba{l}
\psi^+_je^{i(-)^{j}f^+(\lambda)}=\psi^-_je^{i(-)^{j}f^-(\lambda)}=E_j(\lambda),~~j=1,2, 
\ea
\eeq
where the analytic functions $f^{\pm}(\lambda)$, defined by
\beq
\ba{l}
f^{\pm}(\lambda)\equiv \frac{1}{2\pi i}
\int_{\RR}\frac{d\lambda'}{\lambda'-(\lambda\pm i0)}f(W(\lambda')),
\ea
\eeq
exhibit the following asymptotics
\beq\label{asympt_fpm}
\ba{l}
f^{\pm}(\lambda)\sim i\sum\limits_{n\ge 1}\langle \lambda^{n-1}f\rangle \lambda^{-n}, \\
\langle \lambda^n f\rangle \equiv \frac{1}{2\pi}\int_{\RR}\lambda^n(W(\lambda))d\lambda ,
\ea
\eeq
if $f$ decays faster than any power. It follows that
\beq
\ba{l}
E_1(\lambda)\equiv -t\lambda^2-(y+t<f>)\lambda +x-2ut-y<f>- \\
t(<\lambda f>+\frac{1}{2}<f>^2),~~E_2(\lambda)\equiv \lambda -<f>,
\ea
\eeq
implying the following explicit solution of the NRH problem:
\beq\label{sol_dKP}
\ba{l}
\psi^{\pm}_j=E_j(\lambda)e^{i(-)^{j+1}f^+(\lambda)},~~j=1,2.
\ea
\eeq
A characterization of the corresponding solutions of dKP is obtained observing that, 
since $W$ in (\ref{W_prod}) depends on the unknowns $u$ and  $\partial^{-1}_xu_{y}$, isolating the 
$1/\lambda$ term of equations (\ref{sol_dKP}) for $|\lambda |\gg 1$ and comparing them with 
the asymptotics (\ref{psi_asympt_1}),(\ref{psi_asympt_2}), we obtain the algebraic system 
\beq\label{sol_sim}
\ba{l}
q^{(1)}_1=2t\partial^{-1}_xu_{y}-yu=-(x-2ut)<f>+y(<f>^2/2+<\lambda f>+ \\
t(<\lambda^2 f>+<f><\lambda f>+<f>^3/6)), \\
q^{(1)}_2=u=<\lambda f>-<f>^2/2.
\ea
\eeq
for the unknowns $u$ and  $\partial^{-1}_xu_{y}$. The constructed solutions of dKP correspond to the following 
differential reduction (\ref{diff_constr}):  
\beq
3t u_t+xu_x+2yu_y+2u=0.
\eeq 
Substituting its general solution
\beq
u=t^{-2/3}A\left(x',y'\right),~~x'=\frac{x}{t^{1/3}},~~y'=\frac{y}{t^{2/3}},
\eeq
into dKP, one obtains the similarity reduction of dKP:
\beq\label{self_sim_equ}
A_{x'}+\frac{x'}{3}A_{x'x'}+\frac{2}{3}y' A_{x'y'}-A_{y'y'}-(AA_{x'})_{x'}=0.
\eeq 
Therefore the algebraic system (\ref{sol_sim}) characterizes these similarity solutions of dKP.

\subsection{Example 2. The invariant $\psi^+_1 +a \psi^+_2$}
If $S(q,p)=q+a p$, where $a$ is a real parameter, then equation (\ref{sol_x}) read 
\beq
q(t)=q_0+a {\cal H'}(S)(t-t_0),~~p(t)=p_0-{\cal H'}(S)(t-t_0),
\eeq 
becoming the NRH problem
\beq\label{RHsum}
\ba{l}
\psi^+_1=\psi^-_1+iaf(\psi^-_1+a\psi^-_2), \\
\psi^+_2=\psi^-_2- if(\psi^-_1+a\psi^-_2),~~\lambda\in \RR ,
\ea
\eeq
satisfying the symplectic (\ref{dKP}) and reality (\ref{reality1}) constraints.  

Due to the invariance equation (\ref{invariance})
\beq\label{W1}
\psi^+_1+a\psi^+_2=\psi^-_1+a\psi^-_2=-t\lambda^2-(y-a)\lambda+x-2ut \equiv W(\lambda),
\eeq
the NRH problem linearizes and decouples:
\beq
\psi^+_1=\psi^-_1+i a f(W),~~\psi^+_2=\psi^-_2-i f(W),~~\lambda\in\RR,
\eeq
and equations (\ref{sol-RH3}) become 
\beq
\ba{l}
\psi^+_1(\lambda)-iaf^+(\lambda)=\psi^-_1-iaf^-(\lambda)=E_1(\lambda), \\
\psi^+_2(\lambda)+if^+(\lambda)=\psi^-_2+if^-(\lambda)=E_2(\lambda) , \\
E_1(\lambda)=-t\lambda^2-y\lambda+x-2ut,~~ E_2(\lambda)=\lambda .
\ea
\eeq
Then the solution of the NRH problem reads:
\beq\label{sol2_psi1}
\ba{l}
\psi^{\pm}_1=-t\lambda^2-y\lambda+x-2ut+iaf^{\pm}(\lambda),
\ea
\eeq
\beq\label{sol2_psi2}
\ba{l}
\psi^{\pm}_2=\lambda-if^{\pm}(\lambda).
\ea
\eeq

Using equations (\ref{sol2_psi1}),(\ref{sol2_psi2}) and (\ref{psi_asympt_1}), 
it is possible to express the coefficients ${q}^{(n)}_{1,2}$ of the asymptotic 
expansions  in terms of the spectral function $f(W)$ in the following way:
\beq\label{closure1}
\ba{l} 
{q}^{(n)}_1=-a <\lambda^{n-1} f>,~~n\ge 1,
\ea
\eeq
\beq\label{closure2}
\ba{l} 
{q}^{(n)}_2=<\lambda^{n-1} f>,~~n\ge 1.
\ea
\eeq
Since $W$ in (\ref{W1}) is function of ${q}^{(1)}_2=u$ only, equation (\ref{closure2}) for $n=1$ 
\beq\label{sol_exact0}
u=\frac{1}{\sqrt{t}}F\left(x+\frac{(y-a)^2}{4t}-2ut\right). 
\eeq
characterizes the family of solutions associated with the above NRH problem, where 
\beq\label{def_F}
F(z)=\frac{1}{2\pi}\int\limits_{\RR}f(-\mu^2+z)d\mu . 
\eeq
Equation (\ref{sol_exact0}) describes a family of solutions of dKP, constant on the parabola
\beq
x+\frac{(y-a)^2}{4t}=\xi ,
\eeq 
and breaking simultaneously in all points of it. Therefore the above exact solutions 
do not describe the breaking of a localized wave in the $(x,y)$ plane, typical of the 
dKP evolution of localized waves \cite{MS5}. In the next subsection  we show 
how to generalize this solvable case to construct exact solutions 
describing the breaking of a localized wave in a point of the $(x,y)$ plane. 

The corresponding differential constraint (\ref{diff_constr}) reads
\beq\label{diff_red_sum}
(y-a)u_x-2tu_y=0.
\eeq
Indeed, substituting the general solution 
\beq
u=\varphi(\xi,t),~~\xi=x+\frac{(y-a)^2}{4t},
\eeq
of (\ref{diff_red_sum}) into the dKP equation, one obtains the equation $\varphi_t+\varphi/(2t)+\varphi\varphi_{\xi}=0$; 
at last, setting $\varphi=t^{-1/2}v(\xi,\tau),~~\tau=2\sqrt{t}$, we finally derive the Burgers - Hopf equation 
$v_{\tau}+vv_{\xi}=0$, whose general solution reads $v=F(\xi-v\tau)$, where $F$ is an arbitrary function of a single argument. 
Going back to the original variables, we recover (\ref{sol_exact0}). 

\subsection{Example 3. The invariant $\psi^+_1 +a (\psi^+_2)^n$}
If $S(q,p)=q+a p^n$, where $a$ is a real parameter, then equation (\ref{sol_x}) read 
\beq
q(t)=q_0+a p^n_0-a\left(p_0-{\cal H}'(S)(t-t_0)\right)^n ,~~p(t)=p_0-{\cal H'}(S)(t-t_0),
\eeq 
becoming the NRH problem
\beq\label{RHsum}
\ba{l}
\psi^+_1=\psi^-_1+a {\psi^-_2}^n-a\left(\psi^-_2-if(\psi^-_1+a {\psi^-_2}^n)\right)^n, \\
\psi^+_2=\psi^-_2- if(\psi^-_1+a{\psi^-_2}^n),~~\lambda\in \RR ,
\ea
\eeq
whose data satisfy the symplectic (\ref{dKP}) and reality (\ref{reality1}) constraints. 

Due to the invariance equation (\ref{invariance})
\beq\label{W1_n}
\psi^+_1+a{\psi^+_2}^n=\psi^-_1+a{\psi^-_2}^n=
-t\lambda^2-y\lambda+x-2ut +a\left(\left({\psi^-_2}\right)^n\right)_{+}\equiv W(\lambda),
\eeq
the NRH problem linearizes and decouples and equations (\ref{sol-RH3}) become (\ref{W1_n}) and  
\beq
\ba{l}
\psi^+_2(\lambda)+if^+(\lambda)=\psi^-_2+if^-(\lambda)=\lambda , 
\ea
\eeq
Then the solution of the NRH problem reads:
\beq\label{sol3_psi1}
\ba{l}
\psi^{\pm}_1=W(\lambda)-a{\psi^{\pm}_2}^n,
\ea
\eeq
\beq\label{sol3_psi2}
\ba{l}
\psi^{\pm}_2=\lambda-if^{\pm}(\lambda).
\ea
\eeq
Since now $W$ depends on the $(n-1)$ unknowns $u,{q}^{(n)}_2,~n=2,\dots,n-1$, the system of $(n-1)$  equations 
(\ref{closure2}), for $n=1,\dots,n-1$, is an algebraic system characterizing a family of implicit solutions of dKP 
parametrized by the arbitrary real function $f$ on a single variable.

The corresponding differential constraint (\ref{diff_constr}) reads, due to (\ref{sub_hierarchy}),
\beq
\ba{l}
\left(\left(\psi^{\pm}_1+a{\psi^{\pm}_2}^n\right)_{-1}\right)_x=yu_x-2tu_y+a n u_{t_n}=0 ,
\ea
\eeq
where $u_{t_n}$ is the $n^{th}$ flow of the dKP sub-hierarchy (\ref{sub_hierarchy}).

For $n=2$, the invariant 
\beq
\psi^{\pm}_1+a{\psi^{\pm}_2}^2= -\tau_2\lambda^2-y\lambda +x-2u\tau_2\equiv W(\lambda)
\eeq
where $\tau_2=t-a$, is equivalent to (\ref{W1}), up to trivial shifts of $y$ and $t$; therefore the corresponding solution of dKP 
is essentially the same as that in \S 3.2.  

Now we show that, if $n>2$ and even, this family of solutions describes the evolution of a localized two-dimensional wave 
breaking at a point of the $(x,y)$ plane. To do it, 
it is convenient to view (\ref{RHsum}) as a NRH problem on the real line in the shifted variable $\mu$: $\mu=\lambda+y/2t,~t\ne 0$,  
and rewrite the asymptotics (\ref{psi_asympt_1}) in terms of $\mu$:
\beq
\label{psi_asympt_mu}
\ba{l}
\psi^{\pm}_1=-t\mu^2+\xi-2t {\tilde q}^{(1)}_2+
\sum\limits_{n\ge 1}\frac{\tilde q^{(n)}_1}{\mu^n},\\
\psi^{\pm}_2=\mu-\eta+\sum\limits_{n\ge 1}\frac{\tilde q^{(n)}_2}{\mu^n},~~|\mu |\gg 1, 
\ea
\eeq
where, for instance:
\beq
\label{def_tilde_q}
\ba{l}
{\tilde q}^{(1)}_1=\partial^{-1}_{\xi}u_{\eta},~~{\tilde q}^{(2)}_1=-\frac{1}{2t}\partial^{-2}_{\xi}u_{\eta\eta}, \\
{\tilde q}^{(1)}_2=u,~~{\tilde q}^{(2)}_2=-\frac{1}{2t}\partial^{-1}_{\xi}u_{\eta},~~
{\tilde q}^{(3)}_2=\partial^{-1}_{\xi}(\frac{u}{2t}+\frac{1}{4t^2}\partial^{-1}_{\xi}u_{\eta\eta}-u u_{\xi})
\ea
\eeq
and the new space variables $\xi$ and $\eta$ are defined by
\beq\label{def_xi}
\xi\equiv x+\frac{y^2}{4t},~~\eta\equiv \frac{y}{2t}.
\eeq
In addition, equation (\ref{closure2}) is replaced by
\beq\label{closure3}
{\tilde q}^{(n)}_2=\langle\mu^{n-1}f\rangle\equiv \frac{1}{2\pi}\int\limits_{\RR}\mu^{n-1}f(\tilde W(\mu))d\mu .
\eeq
where $\tilde W(\mu)\equiv (\psi^{\pm}_1+a{\psi^{\pm}_2}^n)_+$ is now the non negative part of the Laurent expansion, 
in the $\mu$ variable, of the invariant. 
\vskip 5pt
\noindent
If  $n=3$, then   
\beq
\tilde W(\mu)=a\mu^3-\tau_3\mu^2+3a(\eta^2+u)\mu+\xi-a\eta^3-2u\tau_3+3{\tilde q}^{(2)}_2,
\eeq 
where 
\beq\label{def_tau3}
\tau_3=t+3a\eta.
\eeq 
Since $\tilde W$ depends on the two unknowns $u,{\tilde q}^{(2)}_2$, the system of two equations 
(\ref{closure3}), for $n=1,2$, is an algebraic system characterizing a family of implicit solutions of dKP 
parametrized by the arbitrary real function $f$ of a single variable. 

It is remarkable that, in the longtime regime and for $|a|\ll 1$, it is possible to obtain the following 
explicit asymptotic formulas for such a solution:
\beq\label{sol_cubic1}
\ba{l}
u\sim \frac{1}{\sqrt{\tau_3}}F(\xi-a\eta^3-2u\tau_3), \\
\eta=O(1),~~\xi-a\eta^3-2u\tau_3 =O(1),~~t\gg 1,~~|a|\ll 1,
\ea
\eeq    
where $F$ is defined by (\ref{def_F}).

To show it, we first observe that the condition $|a|\ll 1$ implies:
\beq\label{approx1}
\ba{l}
\int_{\RR}f(\tilde W(\mu))d\mu\sim \int_{\RR}[f(\tilde W^{(0)}(\mu))+af'(\tilde W^{(0)}(\mu))\tilde W^{(1)}(\mu)]d\mu= \\
\int_{\RR}[f(\tilde W^{(0)}(\mu))+af'(\tilde W^{(0)}(\mu)){\tilde W^{(1)}}_{ev}(\mu)]d\mu\sim \\
\int_{\RR}f(\tilde W^{(0)}(\mu)+a{\tilde W^{(1)}}_{ev}(\mu))d\mu,
\ea
\eeq
where $\tilde W=\tilde W^{(0)}+a\tilde W^{(1)}$, with
\beq
\ba{l}
\tilde W^{(0)}(\mu)=\left(\psi^{(1)\pm}_1\right)_+=-t\mu^2+\xi-2ut, \\
\tilde W^{(1)}(\mu)=\left({\psi^{(1)\pm}_2}^3\right)_+=\mu^3-3\eta \mu^2+3(\eta^2+u)\mu-\eta^3-6\eta u+3{\tilde q}^{(2)}_2,
\ea
\eeq
and ${\tilde W^{(1)}}_{ev}(\mu)$ is the even part of $\tilde W^{(1)}(\mu)$:
\beq
{\tilde W^{(1)}}_{ev}(\mu)=-3\eta \mu^2-\eta^3-6\eta u+3{\tilde q}^{(2)}_2.
\eeq
Second, the conditions $t\gg 1,~|\eta |=O(1)$ imply that $\tau_3\gg 1$ and suggest the change of integration variable 
$\mu'=\sqrt{\tau_3}\mu$ in (\ref{approx1}). Consequently $u=O(1/\sqrt{\tau_3})=O(1/\sqrt{t})$; the same change of variables in 
the integrals (\ref{closure3}) implies that ${\tilde q}^{(n)}_2=O(t^{-{\frac{n+1}{2}}})$. Therefore $u$ can be neglected 
wrt $\eta$, and the coefficients ${\tilde q}^{(n)}_2,~n\ge 2$ can be neglected with respect to $u$. 

Consequently 
\beq
\ba{l} 
u\sim\frac{1}{2\pi}\int_{\RR}f(\tilde W^{(0)}(\mu)+a{\tilde W^{(1)}}_{ev}(\mu))d\mu\sim\frac{1}{2\pi}\int_{\RR}f(-\tau_3\mu^2+\xi-a\eta^3-2u\tau_3)\\
=\frac{1}{2\pi\sqrt{\tau_3}}\int_{\RR}f(-{\mu'}^2+\xi-a\eta^3-2u\tau_3)d\mu'
\ea
\eeq
and formula (\ref{sol_cubic1}) is derived. Known the first breaking time $\tau_b$ from the well known formula 
\beq\label{tau_b}
\ba{l}
\tau_b=\frac{1}{4{F'(\zeta_b)}^2}=\mbox{min}_{\zeta\in\RR}\frac{1}{4{F'(\zeta)}^2},~~
F'(\zeta_b)<0,
\ea
\eeq
equation (\ref{def_tau3}) implies that, if $a>0$, the first breaking takes place when $t_b=-\infty$ at $y_b=\infty$. 
As we shall see now, to have breaking in a finite point of the $(x,y,t)$ space, we need to consider $n$ even.

\vskip 5pt
\noindent
If $n=4$, then    
\beq
\ba{l}
\tilde W(\mu)=-t \mu^2+\xi -2ut+a\left({\psi^{\pm}_2}^4\right)_+= \\
-t\mu^2++a[\mu^4-4\eta\mu^3]-(\tau_4-4u)\mu^2+\\
a(-4\eta^3-8\eta u+4 {\tilde q}^{(2)}_2)\mu+ \xi -2u\tau_4 +a[(\eta^4-8\eta {\tilde q}^{(2)}_2+
4{\tilde q}^{(3)}_2+6 u^2)],
\ea
\eeq
where 
\beq\label{def_tau_4}
\tau_4=t-6a\eta^2.
\eeq 
Since now $\tilde W$ depends on the three unknowns $u,{\tilde q}^{(n)}_2,~n=2,3$, the system of three equations 
(\ref{closure3}), for $n=1,2,3$, is an algebraic system characterizing a family of implicit solutions of dKP 
parametrized by the arbitrary real function $f$ on a single variable. 

Following the same derivation as in the previous example, one obtains the following simple one dimensional implicit 
asymptotic formula for such a family of solutions 
\beq\label{sol_quartic1}
\ba{l}
u\sim \frac{1}{\sqrt{\tau_4}}F(\xi+a\eta^4-2u\tau_4), \\
\eta=O(1),~~\xi+a\eta^4-2u\tau_4 =O(1),~~t\gg 1,~~0<a\ll 1,
\ea
\eeq    
where $F$ is again defined by (\ref{def_F}).

Known the first breaking time $\tau_b$ from (\ref{tau_b}), equation (\ref{def_tau_4}) 
implies that, if $a>0$, the first (physical) breaking time $t_b$ is achieved at $y_b=0~(\eta_b=y_b/2t_b=0)$ and coincides 
with $\tau_b$, while $x_b$ follows from
\beq\label{x_b}
x_b=\zeta_b+2F(\zeta_b)\sqrt{t_b}.
\eeq 
We remark that formula (\ref{sol_quartic1}) is an interesting particular case of the 
generic asymptotic formula (\ref{asympt_dKP_1}) obtained in \cite{MS5}. Summarizing, the first wave breaking takes place 
at $t_b=\tau_b$, where $\tau_b$ and $\zeta_b$ are 
given in (\ref{tau_b}), at the point $(\zeta_b+2F(\zeta_b)\sqrt{t_b},0)$ of the $(x,y)$ plane. 

Now we are ready to write the case of an arbitrary power $n$. In this case 
\beq
\tilde W(\mu)=-t \mu^2+\xi -2ut+a\left({\psi^{\pm}_2}^n\right)_+.
\eeq
Since now $\tilde W$ depends on the $(n-1)$ unknowns $u,{\tilde q}^{(n)}_2,~n=2,\dots,n-1$, the system of $(n-1)$  equations 
(\ref{closure3}), for $n=1,\dots,n-1$, is an algebraic system characterizing a family of implicit solutions of dKP 
parametrized by the arbitrary real function $f$ on a single variable. 

Proceeding as in the previous examples, one obtains the following simple one dimensional implicit 
asymptotic formula for such a family of solutions 
\beq\label{sol_n_1}
\ba{l}
u\sim \frac{1}{\sqrt{\tau_n}}F(\xi+(-1)^n a\eta^n-2u\tau_n), \\
\eta=O(1),~~\xi+(-1)^n a\eta^n-2u\tau_n =O(1),~~t\gg 1,~~0<a\ll 1,
\ea
\eeq
where
\beq\label{def_tau_n}
\tau_n=t-(-1)^n a{n\choose n-2}\eta^{n-2}
\eeq    
and $F$ is defined by (\ref{def_F}). Following the same considerations made in the particular case $n=4$, if $n$ is even and $a>0$, 
the localized solution breaks at time $t_b=\tau_b$, where $\tau_b$ is defined in (\ref{tau_b}), in the point   
$(x_b,0)$, where $x_b$ is defined in (\ref{x_b}). 
We remark that also formula (\ref{sol_n_1}) is an interesting particular case of the 
generic asymptotic formula (\ref{asympt_dKP_1}) obtained in \cite{MS5}. 

\section{Solvable NRH problems for the heavenly equation}
In this section we consider two examples of solvable NRH problems allowing to construct two 
different classes of similarity solutions of the heavenly equation.

\subsection{Example 1. The invariant $\psi^+_1 \psi^+_2$ and similarity solutions \cite{MS6}}
If $S(q,p)=q p$, then, as in \S 3.1, the NRH problem reads
\beq
\psi^+_1=\psi^-_1e^{if(\psi^-_1\psi^-_2,\lambda)},~~\psi^+_2=\psi^-_1e^{-if(\psi^-_1\psi^-_2,\lambda)}
\eeq
satisfying the symplectic and reality constraints, where now $f$ is an arbitrary real function of two variables. Then 
the invariance equation (\ref{invariance}) becomes:
\beq\label{W_prod_heav}
\psi^+_1 \psi^+_2=\psi^-_1 \psi^-_2=zt\lambda^2-(xt+yz)\lambda+xy-z\theta_x+t\theta_y \equiv W(\lambda),
\eeq
and the NRH problem linearizes and decouples:
\beq
\ba{l}
\psi^+_1=\psi^-_1e^{if(W(\lambda),\lambda)},~~
\psi^+_2=\psi^-_2e^{-if(W(\lambda),\lambda)}.
\ea
\eeq
Equations (\ref{sol-RH3}) become 
\beq
\ba{l}
\psi^+_je^{i(-)^{j}f^+(\lambda)}=\psi^-_je^{i(-)^{j}f^-(\lambda)}=E_j(\lambda),~~j=1,2, 
\ea
\eeq
where
\beq
\ba{l}
f^{\pm}(\lambda)\equiv \frac{1}{2\pi i}
\int_{\RR}\frac{d\lambda'}{\lambda'-(\lambda\pm i0)}f(W(\lambda'),\lambda').
\ea
\eeq
It follows that
\beq
\ba{l}
E_1(\lambda)\equiv x-z\lambda -z<f> \\
E_2(\lambda)\equiv y-t\lambda +t<f>,
\ea
\eeq
implying the following explicit solution of the NRH problem:
\beq\label{sol1_heav}
\ba{l}
\psi^{\pm}_j=E_j(\lambda)e^{i(-)^{j+1}f^+(\lambda)},~~j=1,2.
\ea
\eeq
A characterization of the corresponding solutions of dKP is obtained observing that, 
since $W$ in (\ref{W_prod_heav}) depends on the unknowns $\theta_x,\theta_y$, isolating the 
$1/\lambda$ terms of equations (\ref{sol1_heav}) for $|\lambda |\gg 1$ and comparing them with 
the asymptotics (\ref{psi_asympt_heav}), we obtain the algebraic system 
\beq\label{sol_sim_heav}
\ba{l}
\theta_y=x<f>-z\left(<\lambda f>+\frac{1}{2}<f>^2\right), \\
\theta_x=y<f>-t\left(<\lambda f>-\frac{1}{2}<f>^2\right) 
\ea
\eeq
for the unknowns $\theta_x,\theta_y$. The constructed solutions of heavenly, parametrized by an arbitrary 
real function of two arguments, correspond to the following differential reduction:  
\beq
-W_{-1}=t\theta_t+y\theta_y-x\theta_x-z\theta_z=0.
\eeq 
Substituting its general solution
\beq
\theta=A(\tilde x,\tilde y,\tilde z),~~\tilde x=xt,~~\tilde y=y/t,~~\tilde z=zt,
\eeq
into the heavenly equation, one obtains the following similarity reduction of (\ref{heavenly}):
\beq\label{self_sim_equ_heav}
\tilde x A_{\tilde x\tilde x}-\tilde y A_{\tilde x\tilde y}+\tilde z A_{\tilde x\tilde z}-A_{\tilde y\tilde z}+
A_{\tilde x\tilde x}A_{\tilde y\tilde y}-A^2_{\tilde x\tilde y}   =0.
\eeq 
Therefore the algebraic system (\ref{sol_sim_heav}) characterizes the above similarity solutions of heavenly.

\subsection{Example 2. The invariant $(\psi^+_1)^2 +(\psi^+_2)^2$ and rotationally invariant solutions }
If $S(q,p)=(q^2+p^2)/2$, then equation (\ref{sol_x}) reads 
\beq
\ba{l}
q(t)=\cos{\cal H'}(S)(t-t_0)~ q_0+\sin{\cal H'}(S)(t-t_0)~ p_0,  \\
p(t)=-\sin{\cal H'}(S)(t-t_0) ~q_0+\cos{\cal H'}(S)(t-t_0) ~p_0,
\ea
\eeq 
corresponding to the NRH problem
\beq
\ba{l}
\psi^+_1=\cosh f\left(\frac{{\psi^-_1}^2+{\psi^-_2}^2}{2},\lambda\right)~\psi^-_1+
i\sinh f\left(\frac{{\psi^-_1}^2+{\psi^-_2}^2}{2},\lambda\right)~\psi^-_2 , \\ 
\psi^+_2=-i\sinh f\left(\frac{{\psi^-_1}^2+{\psi^-_2}^2}{2},\lambda\right)~\psi^-_1+
\cosh f\left(\frac{{\psi^-_1}^2+{\psi^-_2}^2}{2},\lambda\right)~\psi^-_2 
\ea
\eeq
satisfying the symplectic and reality constraints. Then the invariance equation (\ref{invariance}) becomes: 

\beq
\ba{l}
\frac{{\psi^+_1}^2+{\psi^+_2}^2}{2}=\frac{{\psi^-_1}^2+{\psi^-_2}^2}{2}=
\frac{z^2+a t^2}{2}\lambda^2-(xz+aty)\lambda+\frac{x^2+a y^2}{2}+ \\
z\theta_y-at\theta_x \equiv W(\lambda),
\ea
\eeq
and the NRH problem linearizes:
\beq
\ba{l}
\psi^+_1=\cosh\left(f(W,\lambda)\right)~\psi^-_1-i\sinh\left(f(W,\lambda)\right)~\psi^-_2 , \\ 
\psi^+_2=i\sinh\left(f(W,\lambda)\right)~\psi^-_1+\cosh\left(f(W,\lambda)\right)~\psi^-_2 . 
\ea
\eeq 
Equations (\ref{sol-RH3}) become
\beq
\ba{l}
\cosh f^{\pm}(\lambda)\psi^{\pm}_1-i\sinh f^{\pm}(\lambda)\psi^{\pm}_2=E_1(\lambda), \\
i\sinh f^{\pm}(\lambda)\psi^{\pm}_1+\cosh f^{\pm}(\lambda)\psi^{\pm}_2=E_2(\lambda),
\ea
\eeq
where
\beq
\ba{l}
E_1(\lambda)\equiv -z\lambda +x-t\langle f\rangle, \\
E_2(\lambda)\equiv -t\lambda +y+z\langle f\rangle.
\ea
\eeq
Therefore the explicit solution of the NRH problem reads
\beq\label{sol2_heav}
\ba{l}
\psi^{\pm}_1(\lambda)=\cosh f^{\pm}(\lambda) E_1(\lambda)+i\sinh f^{\pm}(\lambda) E_2(\lambda), \\
\psi^{\pm}_2(\lambda)=-i \sinh f^{\pm}(\lambda) E_1(\lambda)+\cosh f^{\pm}(\lambda) E_2(\lambda), 
\ea
\eeq
A characterization of the corresponding solutions of heavenly is obtained observing that, 
since $W$ in (\ref{W_prod_heav}) depends on the unknowns $\theta_x,\theta_y$, isolating the 
$1/\lambda$ terms of equations (\ref{sol1_heav}) for $|\lambda |\gg 1$ and comparing them with 
the asymptotics (\ref{psi_asympt_heav}), we obtain the algebraic system 
\beq\label{sol2_sim_heav}
\ba{l}
\theta_y=y<f>-t\left(<\lambda f>-\frac{z}{2}<f>^2\right), \\
\theta_x=x<f>-z\left(<\lambda f>-\frac{t}{2}<f>^2\right) 
\ea
\eeq
for the unknowns $\theta_x,\theta_y$. These solutions, parametrized by an arbitrary 
real function of two arguments, correspond to the following differential reduction:  
\beq
-W_{-1}/2=z\theta_t+x\theta_y-a y \theta_x-a t \theta_z=0.
\eeq 
Substituting its general solution 
\beq
\theta=B(\alpha,\beta,\gamma),~~\alpha=x^2+a y^2,~~\beta=z^2+a t^2,~~\gamma=xt-yz,
\eeq
into the heavenly equation, one obtains the following reduction of (\ref{heavenly}):
\beq\label{self_sim_equ2_heav}
\ba{l}
2B_{\gamma}+\gamma B_{\gamma\gamma}+2\alpha B_{\alpha\gamma} +2\beta B_{\beta\gamma} +4 a\gamma B_{\alpha\beta} + 
2 B_{\alpha}(\beta B_{\gamma\gamma} +2aB_{\alpha} + \\ 4a\alpha B_{\alpha\alpha} +4a\gamma B_{\alpha\gamma}) +
4(\alpha\beta-\gamma^2)(B_{\alpha\alpha}B_{\gamma\gamma}-B^2_{\alpha\gamma})=0.
\ea
\eeq 
Therefore the algebraic system (\ref{sol2_sim_heav}) characterizes a class of rotationally invariant solutions of heavenly.

\vskip 5pt
\noindent
{\bf Acknowledgements}. This research has been supported by the RFBR 
grants 07-01-00446, 08-01-90104, and 09-01-92439, by the bilateral agreement between the Consortium Einstein 
and the RFBR, and by the bilateral agreement between the 
University of Roma ``La Sapienza'' and the Landau Institute for Theoretical Physics of the 
Russian Academy of Sciences. 
 

\end{document}